\documentclass[12pt,preprint]{emulateapj}

\usepackage{amssymb}
\usepackage{times}
\usepackage{amsmath}
\usepackage{subfigure}
\usepackage{multirow}
\usepackage[colorlinks,dvipdfm,linkcolor=blue,anchorcolor=,citecolor=blue]{hyperref}
\makeatletter

\newcommand{\teff}{$T_{\mathrm{eff}}$}

\newcommand{\kepler}{\textit{Kepler}}

\newcommand{\keplermission}{\textit{Kepler Mission}}

\newcommand{\msol}{M$_\sun$}

\newcommand{\rmnum}[1]{\romannumeral #1}
\newcommand{\Rmnum}[1]{\expandafter\@slowromancap\romannumeral #1@}
\makeatother

\slugcomment{}

\shorttitle{\textit{Asteroseismic Analyzing of {\it KIC} 6225718}}
\shortauthors{\textit{Wu \& Li}}

\begin{document}


\title{New Insights for High-precision Asteroseismology: Acoustic Radius of {\it KIC} 6225718}


\author{
Tao Wu\altaffilmark{1,2} and Yan Li \altaffilmark{1,2}
}


\altaffiltext{1}{Yunnan Observatories, Chinese Academy of Sciences, P.O. Box 110, Kunming 650011, China; wutao@ynao.ac.cn, ly@ynao.ac.cn}
\altaffiltext{2}{Key Laboratory for Structure and Evolution of Celestial Objects, Chinese Academy of Sciences, P.O. Box 110, Kunming 650011, China}


\begin{abstract}
  Asteroseismology is a powerful tool for probing stellar interiors and determining stellar fundamental parameters. In previous works, $\chi^2$-minimization method is usually used to find the best matching model to characterize observations. In this letter, we adopt the $\chi^2$-minimization method but only using the observed high-precision oscillation to constrain theoretical models for solar-like oscillating star KIC 6225718, which is observed by \kepler\ satellite. We also take into account the influence of model precision. Finally, we find that the time resolution of stellar evolution can not be ignored in high-precision asteroseismic analysis. Based on this, we find the acoustic radius $\tau_{0}$ is the only global parameter that can be accurately measured by $\chi^2_{\nu}$ matching method between observed frequencies and theoretical model calculations. We obtain $\tau_{0}=4601.5^{+4.4}_{-8.3}$ seconds. In addition, we analyze the distribution of $\chi^2_{\nu}$-minimization models (CMMs), and find that the distribution range of CMMs is slightly enlarged by some extreme cases, which possess both of larger mass and higher (or lower) heavy element abundance, at lower acoustic radius end.

\end{abstract}



\section{Introduction}\label{sec-intr}
Asteroseismology is a very effective tool to determine stellar fundamental parameters, such as stellar mass, radius, and surface gravity. It is also a direct way to explore the internal structure and evolutionary status of stars when combined with stellar theoretical model. It has already been used as well to determine cluster fundamental parameters \citep[such as][]{basu11,Miglio12,Balona13,Wu14a,Wu14b}. Thanks to the space based observation space projects, such as {\it CoRoT} \citep{Baglin06}, \kepler\ \citep{Borucki2010Sci}, and the following K2 mission \citep[][]{Howell2014PASP}, more and more ultraprecise and long-term photometric time series have been obtained. They have revolutionized asteroseismology in many aspects, such as investigating the internal differential rotation of stars \citep[e.g.][]{Beck2012Natur,Mosser2012AAb,Deheuvels2012ApJ,Deheuvels2014AA,Goupil2013AA} and probing the properties of the convective zone and overshooting \citep[e.g.][]{Soriano2008CoAst,Deheuvels2011AA,Lebreton2012AA,Guenther2014ApJ,Tian2014MNRAS,Moravveji2015AA}.

The ultimate goal of asteroseismology is to reveal the internal structure and evolutionary status of stars. To probe the internal structure or a specific physical process in a star, however, we need to compare stellar theoretical models with observations of the star \citep[e.g.][]{Guenther2004ApJ,Deheuvels2012ApJ,Deheuvels2014AA,Lebreton2012ASPC,Lebreton2014AA,
Tian2014MNRAS}. Usually, the goodness of the matching between theoretical models and observations is evaluated through a $\chi^2$ method, which is
defined as
\begin{equation}\label{eq_chi2}
\chi^2_{x}=\frac{1}{N}\sum^{N}_{i=1}\left(\frac{x_{i}^{\rm obs}-x_{i}^{\rm mod}}{\sigma_{x_{i}^{\rm obs}}}\right)^{2}
\end{equation}
\citep{Eggenberger2004AA}, where the superscript {\bf\footnotesize obs} and {\bf\footnotesize mod} represent the observations and theoretical model calculations of a quantity $x_{i}$, respectively. Correspondingly, $\sigma_{x_{i}^{\rm obs}}$ denotes the uncertainty of observed quantity $x_{i}^{\rm obs}$ and $N$ is the total number of available observations of $x_{i}^{\rm obs}$. Obviously, a model with the minimum $\chi^2_{x}$ is the best matching model.

To obtain the best matching model, asteroseismic parameters (mainly frequencies) and non-asteroseismic parameters (i.e. parameters of stellar atmosphere) are usually combined together to get a total residual error ($\chi^2_{x}$) \citep[e.g.][]{Lebreton2012ASPC,Lebreton2014AA}. Sometimes these two kinds of parameters are used separately \citep[e.g.][]{Deheuvels2012ApJ,Deheuvels2014AA,Tian2014MNRAS}. The non-seismic parameters are used as a first constraint to narrow the model parameter range, and then the seismic parameters are used to find the best matching model (i.e., $\chi^2$-minimization model).

Owing to space-based observations (such as {\it CoRoT}, \kepler, and K2), dozens of oscillation frequencies can be obtained for a pulsating star (such as solar-like stars) with extraordinarily high precision. The errors of those observed frequencies are usually much smaller than those of non-asteroseismic parameters (e.g. effective temperature \teff, luminosity $L$, and metal abundance [Fe/H]). However, such high precision of asteroseismic data will be deteriorated by low precision of non-asteroseismic parameters, when they are used together.  Therefore, the precision of the final result will be reduced.

In practice, theoretical models also possess considerable errors. The definition of $\chi^2_{x}$ (Equation \eqref{eq_chi2}) should therefore be modified as:
\begin{equation}\label{eq_chi2-1}
\chi^2_{x}\equiv\frac{1}{N}\sum^{N}_{i=1}\frac{\left(x_{i}^{\rm obs}-x_{i}^{\rm mod}\right)^{2}}{\sigma_{x_{i}^{\rm obs}}^2+\sigma_{x_{i}^{\rm mod}}^2}.
\end{equation}

But what does such matching process tell us about the observed star? Or, in other words, what physical parameters of the observed star can be determined best with such a method? In this letter, we try to answer this question, using only asteroseismic parameters of a solar-like star KIC 6225718.

\section{The precision of stellar models}\label{sec-pre.the.mod}

The theory of stellar structure and evolution tells us that the precision of theoretical models of a star is affected by the space resolution of stellar structure and time resolution of stellar evolution when the input physics and initial parameters are fixed. In this section, we will discuss the influences coming from both of them.

\subsection{Model input physics}\label{sec-input}
We use the Modules of Experiments in Stellar Astrophysics (MESA) evolutionary code, which is developed by \citet{MESA2011}, to calculate theoretical models. It can be used to calculate stellar evolutionary models and their corresponding oscillation information \citep{MESA2013}. We use the package {\small \textbf{``pulse"}} of version {\bf \small ``v6208"} to make our calculations for both stellar evolution and oscillations \citep[]{jcd2008,MESA2011,MESA2013}.

Based on the default parameters, we adopt the OPAL opacity table GS98 \citep{gs98a} series. We choose Eddington grey-atmosphere $T-\tau$ relation as the stellar atmosphere model, and treat the convection zone by the standard mixing-length theory (MLT) of \citet{cox1968} with $\alpha_{\rm MLT}$=2.0. We treat the element diffusion in our model calculations according to \cite{Thoul1994ApJ}. For convective core overshooting, \cite{Tian2014MNRAS} suggested that it is between 0 and 0.2$H_{p}$ and its effect can be ignored. We hence do not consider it in our model calculations. In addition, the semi-convection, thermohaline mixing, and mass loss are not included in this work.

\begin{figure*}
\begin{center}
\includegraphics[scale=0.45,angle=-90]{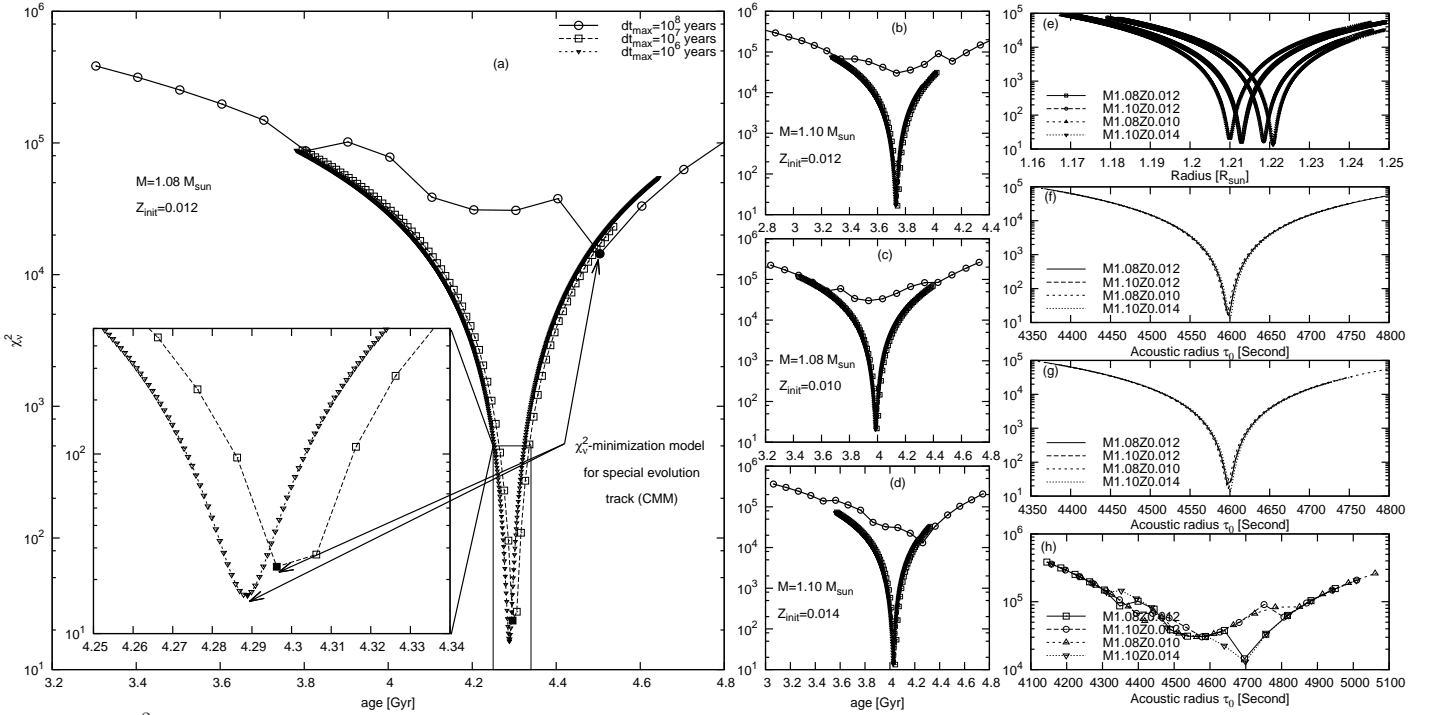}
  \caption{$\chi^{2}_{\nu}$ as a function of stellar age (panels (a)-(d)), stellar radius (panel (e)), and of acoustic radius (panels (f)-(h)). For panels (a)-(d), different panels represent different initial combinations for stellar mass $M$ and metal mass fraction $Z_{\rm init}$. They are (1.08 \msol, 0.012) -- panel (a), (1.10 \msol, 0.012) -- panel (b), (1.08 \msol, 0.010) -- panel (c), and (1.10 \msol, 0.014) -- panel (d), respectively. In panels (a)-(d), different symbols correspond to different time resolutions in stellar evolution. They are $10^{8}$ years -- open circles, $10^{7}$ years -- open squares, and $10^{6}$ years -- open triangles, respectively. In panel (a), the filled points present the $\chi^{2}_{\nu}$-minimization models (CMM) of corresponding time resolution. In addition, the small panel of panel (a) is a zoom. Panels (e) and (f) show the evolutionary track with the time resolution of $10^6$ years. The time resolution of panels (g) and (h) are $10^7$ and $10^8$ years, respectively. The oscillation frequencies $\nu_{i}$ come from \citet{Tian2014MNRAS}. }\label{fig-chi}
\end{center}
\end{figure*}

\subsection{The influence of space and time resolution on stellar models}\label{sec-precision}

For a low-mass main-sequence star (such as KIC 6225718), the radial nodes $n$ of observable oscillation modes are around 20. Moreover, such radial nodes almost distribute over the whole stellar interior. In order to discuss the influence of space resolution on the stellar structure and on its oscillations, we compare two models that consist of 1800 and 3500 shells, respectively. Finally, we find that the discrepancies between them are on the order of 0.04 $\mu$Hz for oscillation frequencies and 0.027 for $\chi^2_{\nu}$, when we match our theoretical models with the observations of KIC 6225718. Therefore, a spacial resolution of about 2000 shells is sufficient to identify accurate stellar models for such a star.

Based on the above input physics, we calculate stellar evolution models and their oscillation frequencies of four initial combinations of mass $M$ and initial heavy element mass fraction $Z_{\rm init}$. They are (1.08 \msol, 0.012), (1.10 \msol, 0.012), (1.08 \msol, 0.010), and (1.10 \msol, 0.014), respectively. In order to check the effect of the time resolution, we assign three different values to the time step of evolution. They are $10^8$, $10^7$, and $10^6$ years, respectively. Finally, we calculate $\chi^2_{\nu}$ value for each model on those tracks with respect to the 33 observed frequencies of KIC 6225718 \citep{Tian2014MNRAS} via Equation \eqref{eq_chi2} and show them in Figure \ref{fig-chi}, where each line corresponds to one evolutionary track. In addition, the $\chi^2_{\nu}$-minimization model of every evolutionary track is also decided.

It can be seen in Figure \ref{fig-chi} that the time resolution of stellar evolution significantly affects the position of the best matching models in the $\chi^2_{\nu}$ diagram. The $\chi^2_{\nu}$-profile with the time step of $10^8$ years show some irregular variations. For shorter time steps, the $\chi^2_{\nu}$-profile become more and more regular and tend to converge.

Panels (a) to (d) in Figure \ref{fig-chi} also show that different time resolutions lead to distributions of $\chi^2_{\nu}$ with great diversity. It should be noted that the $\chi^2_{\nu}$ value indicates how good a model represents the observations. Large diversity and irregular variations of $\chi^2_{\nu}$ thus indicate that the time resolution affects the precision of stellar models. It can be noticed that a too low time resolution in stellar evolution calculations may lead to wrong selection of the best matching model. When the time resolution increases to $10^6$ years, the $\chi^2_{\nu}$-profile begins to converge to a regular ``V'' shape, which defines a unique minimum $\chi^2_{\nu}$. This characteristic removes any ambiguity of how to select the best matching model and ensures that only one model can be selected as the best matching one for a specific evolutionary track. By comparing two models that have time resolutions of $10^6$ and $2\times10^6$ years respectively, we find that there is 0.035 $\mu$Hz for the frequencies, which translates into 0.024 for $\chi^2_{\nu}$. At this moment the precision of theoretical models are higher than that of observations.

According to the stellar oscillation theory, p-mode oscillations are basically acoustic waves. Their properties can be roughly characterized by the acoustic radius $\tau_{0}$, which is defined as
\begin{equation}\label{eq-tau}
\tau_{0}=\int^{R}_{0}\frac{dr}{c_{\rm s}(r)}
\end{equation}
\citep[e.g.][]{Aerts2010}, where $c_{\rm s}$ is the adiabatic sound speed and $R$ is the stellar radius. The acoustic radius $\tau_{0}$ is an important parameter in asteroseismology. For instance, \cite{Ballot2004AA} and \citet{Miglio2010AA} used it as a reference to show the position of the base of convective envelope and the position of helium ionization zone. Equation \eqref{eq-tau} shows that $\tau_{0}$ is determined by stellar radius $R$ and the profile of adiabatic sound speed $c_{\rm s}(r)$. Therefore, such ``V'' shapes in $\chi^2_{\nu}(t_{\rm age})$ may also appear in $\chi^2_{\nu}(R)$ and $\chi^2_{\nu}(\tau_{0})$ (see Figures \ref{fig-chi}(e)-\ref{fig-chi}(h)).

It can be found from Figures \ref{fig-chi}(e) and \ref{fig-chi}(f) that the stellar radius $R$ dominantly determines $\chi^2_{\nu}$ to vary with ``V'' shape, while the profile of sound speed $c_{\rm s}$ fine tunes the minimum of $\chi^2_{\nu}$ to be located at almost the same value of the acoustic radius $\tau_{0}$. In addition, Figures \ref{fig-chi}(f)-\ref{fig-chi}(h) show that  such regular "V" shape gradually disappears when the time step of stellar evolution increases. For low-mass main sequence stars, the order of $10^6$ years for the time resolution is available to obtain accurate enough stellar evolutionary models.

\renewcommand{\arraystretch}{0.75}
\begin{deluxetable}{lcl}
\tabletypesize{\scriptsize}
\tablecaption{Basic parameters of KIC 6225718.\label{table_1} }
\tablehead{
\colhead{Para.} & \colhead{Value} & \colhead{Reference} 
}
\startdata
[Fe/H]         & $-$0.10$\pm$0.12 dex & \cite{Clementini1999MNRAS} \\
               & $-$0.24 dex           & \cite{Nordstrom2004AA}     \\
               & $-$0.19 dex           & \cite{Holmberg2007AA}      \\
               &  $-$0.15 dex         & \cite{Casagrande2011AA} \\
               & $-$0.17$\pm$0.06 dex &  \cite{Bruntt2012MNRAS} \\
               & $-$0.23$\pm$0.15 dex  & \cite{Molenda2013MNRAS}    \\

 $M$          &   1.144$~\rm M_{\sun}$                    & \cite{Boyajian2013ApJ}  \\
              &   $1.08^{+0.07}_{-0.05}~\rm M_{\sun}$     & \cite{Nordstrom2004AA}  \\
              &   $1.17^{+0.04}_{-0.04}~\rm M_{\sun}$    & \cite{Casagrande2011AA} \\
              &   $1.17^{+0.05}_{-0.05}~\rm M_{\sun}$     & \cite{Casagrande2011AA} \\
              &1.31$\pm$0.11$~\rm M_{\sun}$     & \cite{Huber2012ApJ}  \\
              &   1.37$\pm$0.15$~\rm M_{\sun}$  & \cite{Huber2012ApJ}  \\
              &  $1.209^{+0.037}_{-0.034}~\rm M_{\sun}$  & \cite{Silva-Aguirre2012ApJ} \\
              &   $1.17^{+0.04}_{-0.05}~\rm M_{\sun}$  & \cite{Chaplin2014ApJS}   \\
              &   $1.10^{+0.04}_{-0.03}~\rm M_{\sun}$   & \cite{Tian2014MNRAS}     
\enddata
\end{deluxetable}
\renewcommand{\arraystretch}{1}

\section{Observations of KIC 6225718}\label{sec-data}
\subsection{KIC 6225718}
KIC 6225718 is also named as HIP (HIC) 97527 and HD 187637. It is a low-mass main-sequence star \citep{Boyajian2013ApJ}. It has been included in various studies. We summarieze attempts to determine its mass and metal abundance in Table \ref{table_1}.

Since it had been observed by \kepler, \cite{Bruntt2012MNRAS}, \cite{Silva-Aguirre2012ApJ}, \cite{Huber2012ApJ}, \cite{Chaplin2014ApJS}, and \cite{Tian2014MNRAS} investigated it with different methods. They analyzed the oscillation power spectrum and extracted $\nu_{\rm max}=2301-2352~\mu$Hz and $\Delta\nu\simeq105.8~\mu$Hz. \cite{Bruntt2012MNRAS}, \cite{Silva-Aguirre2012ApJ}, \cite{Huber2012ApJ}, and \cite{Chaplin2014ApJS} determined its fundamental parameters by combining these asteroseismic parameters with other non-asteroseismic parameters. \cite{Tian2014MNRAS} extracted 33 individual frequencies from its oscillation spectrum. The radial order $n$ of the identified oscillation modes ranges from 13 to 24 and the spherical harmonic degree $l=0,~1,\rm and ~2$. They used the ratio $r_{01}=\frac{\delta\nu_{01}(n)}{\Delta\nu_{l=1}(n)}$ to analyze the overshooting of convective core and found that it is about 0-0.2$H_{\rm p}$.

\subsection{Observations}\label{sec-obs}
KIC 6225718 was observed with \kepler\ in two different observing modes: short- (about 1 minute exposure time) and long-cadence (about 30 minutes exposure times). There are 18 quarters of long-cadence (Q0--Q17) and 13 quarters of short-cadence (Q1 and Q6--Q17) photometric data available. In this letter, we use the 33 individual frequencies and their uncertainties from \citep[see Table 2 of][]{Tian2014MNRAS}. For more details on the data processing we refer to Sec. 2 in \citet[][]{Tian2014MNRAS}.

\begin{figure}
\begin{center}
\includegraphics[scale=0.49,angle=-90]{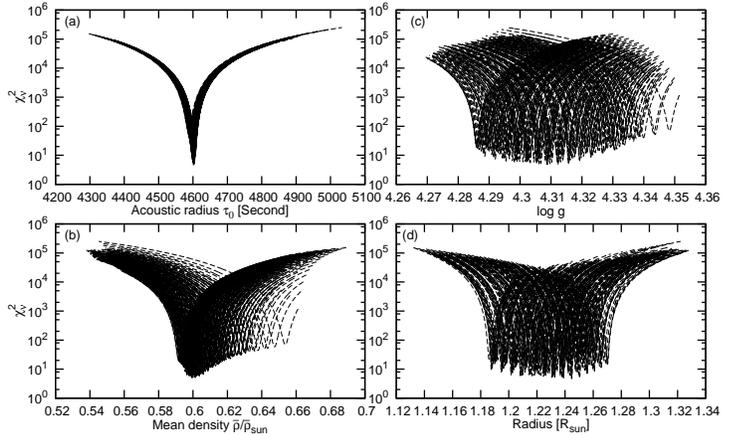}
\caption{$\chi^2_{\nu}$ as a function of acoustic radius $\tau_{0}$ (panel (a)), mean density $\bar\rho$ (panel (b)), surface gravity $\log g$ (panel (c)), and radius $R$ (panel (d)), respectively, for all of the 196 calculated evolution tracks. }\label{fig.tau.evol}
\end{center}
\end{figure}

\begin{figure}
\begin{center}
\includegraphics[scale=0.49,angle=-90]{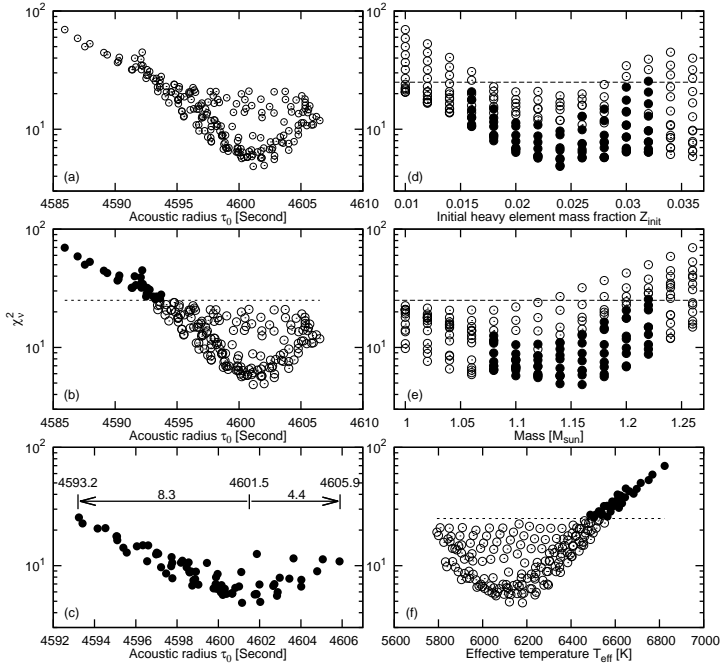}
\caption{$\chi^2_{\nu,\rm CMM}$ as a function of acoustic radius $\tau_{0}$ (panels (a)-(c)), initial metal mass fraction $Z_{\rm init}$ (panel (d)), mass $M$ (panel (e)), and of effective temperature $T_{\rm eff}$ (panel (f)), respectively. Panel (a) shows all of calculated CMMs. Panel (b) is same with panel (a), but it is divided into two parts by dashed line ($\chi^2_{\nu,\rm CMM}=25$). Panel (c) presents CMMs $M\in[1.08,1.22]{\rm M_{\odot}}$ and $Z_{\rm init}\in[0.016,0.032]$. In panels (d) and (e), filled points correspond to panel (c) and dashed line correspond to panel (b). Panel (f) is similar to panel (b), but varies with effective temperature $T_{\rm eff}$. }\label{fig.tau.CMM}
\end{center}
\end{figure}

\section{Calculations and results}\label{sec-modelcal.resul}

\subsection{ Modeling and $\chi^2_{\nu}$ matching}\label{sec-calibration}
According to Table \ref{table_1} and preliminary calculations, we set the initial stellar mass $M$ ranging from 1.0 \msol\ to 1.26 \msol\ with steps of 0.02 \msol\ and the initial heavy element mass fraction $Z_{\rm init}$ from 0.010 to 0.036 with steps of 0.002, respectively. Based on above input physics (see Sec. \ref{sec-input}), we calculate a total of 196 evolutionary tracks. For each model on these tracks we compute the p-mode spectrum and compare it to the observed frequencies. The resulting $\chi^2_{\nu}$ values are shown in Figure \ref{fig.tau.evol} as a function of various model parameters.

As already shown in Section \ref{sec-precision}, each evolutionary track has one and only one $\chi^2_{\nu}$-minimization model. Hence we call such models as $\chi^{2}_{\nu}$-minimization model for the considered evolutionary tracks (hereafter CMM). They are indicated as filled symbols in Figure \ref{fig-chi}(a). The CMMs of all evolutionary tracks are shown in Figure \ref{fig.tau.CMM}.

\subsection{Precise determination of acoustic radius $\tau_{0}$ }\label{sec-res.tau}

Figure \ref{fig.tau.evol}(a) indicates that $\chi^2_{\nu}$ profiles of all calculated evolutionary tracks almost overlap with each other when they are plotted against the acoustic radius $\chi^2_{\nu}(\tau_{0})$. They individual CMMs all have an acoustic radius of $\tau_{0}\approx4600$ seconds. On the other hand, the $\chi^2_{\nu}$-profiles do not overlap when plotted as function of other model parameters (see Figure \ref{fig.tau.evol}(b)-\ref{fig.tau.evol}(d)) but cover a rather large range of the given parameters. From this we conclude that the acoustic radius is global model parameter that can be determined best from matching observed frequencies with theoretical ones.

Figure \ref{fig.tau.CMM}(a) shows that the distribution of CMMs ($\chi^2_{\nu,\rm CMM}(\tau_{0})$) looks like a ladle. Most of the models are located in the bowl-shaped part and the rest form the handle. In Figure \ref{fig.tau.CMM}(b), those CMMs are divided into two parts by a dashed line of $\chi^2_{\nu,\rm CMM}=25$. One part is the bowl-shaped part ($\chi^2_{\nu,\rm CMM}<25$), which is shown with open circles and referred to as normal part. The other part is the handle part ($\chi^2_{\nu,\rm CMM}>25$), which is denoted with filled circles and referred to as outlying part. It is evident that CMMs in the outlying part have bigger $\chi^2_{\nu}$ than those of the normal part. Those CMMs in the outlying part in the determination of the acoustic radius of KIC 6225718 should be excluded. Figure \ref{fig.tau.CMM}(c) shows those CMMs with $1.08<M/{\rm M_{\odot}}<1.22$ and $0.016<Z_{\rm init}<0.032$. They are all located at the central part of Figures \ref{fig.tau.CMM}(d) and \ref{fig.tau.CMM}(e) and as well showed with filled circles in Figures \ref{fig.tau.CMM}(d) and \ref{fig.tau.CMM}(e). Compared to Figures \ref{fig.tau.CMM}(a) and \ref{fig.tau.CMM}(b), they form almost the bottom edge of the bowl-shaped part. Then the acoustic radius of KIC 6225718 can be evaluated as $4601.5^{+4.4}_{-8.3}$ seconds, where 4601.5 is the position of $\chi^2_{\nu,\rm CMM}$-minimization, and +4.4 and -8.3 are respectively the upper and the lower limits of the profile (see the annotation of Figure \ref{fig.tau.CMM}(c)). Here, our results (Figures \ref{fig.tau.evol}(a), \ref{fig.tau.CMM}(a)-\ref{fig.tau.CMM}(c)) show that the acoustic radius $\tau_{0}$ (of KIC 6225718) can be directly and precisely determined from $\chi^2_{\nu}(\tau_{0})$ without any other extra constraints.

Figures \ref{fig.tau.CMM}(d) and \ref{fig.tau.CMM}(e) show distributions of $\chi^2_{\nu}$ with respect of the initial metal abundance and stellar mass, respectively. The minimum of $\chi^2_{\nu,\rm CMM}$ is realized by stellar models with $Z_{\rm init}=0.024$, and $M=1.14$ \msol\ and 1.16 \msol. It can be seen in Figure \ref{fig.tau.CMM}(d) that CMMs of the normal part occupy the central part, while CMMs of the outlying part are located at both high and low side of the initial metal abundance. Meanwhile, CMMs of the outlying part occupy the higher mass end in Figure \ref{fig.tau.CMM}(e). These facts indicate that parameters such as the stellar mass $M$ and metal abundance $Z_{\rm init}$ may slightly affect the determination of the acoustic radius $\tau_{0}$, and their influences are mainly to enlarge uncertainty on the lower acoustic radius $\tau_{0}$ end.

Figure \ref{fig.tau.CMM}(f) shows that CMMs on the outlying part have higher effective temperature $T_{\rm eff}$ than those of normal part. Their radii $R$ are somewhat larger than normal CMMs', but the increase in $R$ can not fully offset the increase in the profile of sound speed $c_{s}(r)$, i.e., of temperature $T(r)$. Hence those CMMs have smaller acoustic radius $\tau_{0}$ (see Equation \eqref{eq-tau}).

\begin{figure}
\begin{center}
\includegraphics[scale=0.45,angle=-90]{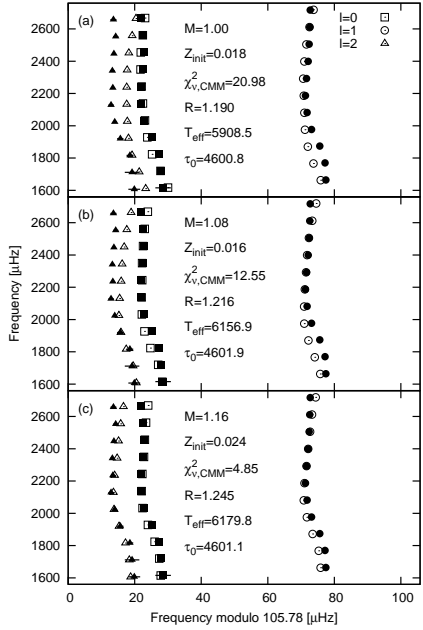}
\caption{\'{E}chelle diagrams of KIC 6225718 (filled points) and corresponding CMMs (open points). Their fundamental parameter are signed with text in figure, e.g., $M$, $Z_{\rm init}$, $\chi^2_{\nu,\rm CMM}$, $R$, $T_{\rm eff}$, and $\tau_{0}$. }\label{fig.echell}
\end{center}
\end{figure}

Figure \ref{fig.echell} (\'{E}chelle diagram) exhibits the goodness of matching between model calculations and observations for three CMMs. Figure \ref{fig.echell}(c) shows results of our best matching model, its parameters are $Z_{\rm init}=0.024$, and $M=1.16$ \msol. Its $\chi^2_{\nu,\rm CMM}=4.58$. For Figures \ref{fig.echell}(a) and \ref{fig.echell}(b), their $\chi^2_{\nu,\rm CMM}$ are close to 21 and 13, respectively. In Figure \ref{fig.echell}(c), the calculated frequencies are in good agreement with the observations of KIC 6225718. In Figures \ref{fig.echell}(a) and \ref{fig.echell}(b), however, our CMMs seem to result in small frequency separations ($\delta\nu_{02}$) that are significantly smaller than that from the observations.

For the surface effect of oscliiations \citep[e.g.,][]{Kjeldsen2008ApJ,Gruberbauer2013MNRAS}, we compare some CMMs that are corrected with those uncorrected, and find that the parameters of CMMs are slightly affected by the surface effect. For example, the discrepancy between them is just about 0.4 seconds in stellar acoustic radius $\tau_{0}$, and about $(1-3)\times10^6$ years in stellar age. This is because the surface effect correctness mainly changes the high-frequency end \citep[e.g.,][]{Gruberbauer2013MNRAS}. While, the value of residual errors between the observations and the CMMs are larger at the part of high-frequency (see Figure \ref{fig.echell}). Hence, the surface effect correctness mainly reduces the value of $\chi^2_{\nu,\rm CMM}$, but only slightly change the fundamental parameters of CMMs. Comparing to the profile of CMMs in $\tau_{0}$ (see Figures \ref{fig.tau.CMM}(a)-\ref{fig.tau.CMM}(c)), we ignore the influence of the surface effect in theoretical models.

\section{Summary and conclusions}\label{sec-sc}
The goal of asteroseismology is to probe stellar interiors and to determine stellar fundamental parameters. In this letter, we only use the high-precision asteroseismic observations (i.e., oscillation frequencies) to constrain theoretical model to analyze the solar-like oscillation star KIC 6225718. On the other hand, we consider the influence of the model uncertainties for an asteroseismic analysis. For this investigation we conclude the following:

\rmnum{1}: The time resolution of stellar evolution seriously affects stellar evolution and structure, as well the corresponding oscillations. Hence it can not be ignored in high-precision asteroseismology.

\rmnum{2}: The stellar acoustic radius is an important parameter in characterizing the properties of acoustic oscillations. The calculations suggest that it is the global parameter that can be measured best by $\chi^2_{\nu}$ matching method between observed frequencies and theoretical model calculations. The calculations also suggest that the distribution range of CMMs is slightly enlarged for some extreme cases at the end of lower acoustic radius. They have both of higher stellar mass and smaller (or larger) heavy element abundance.

\rmnum{3}: Finally, we obtain the acoustic radius $\tau_{0}=4601.5^{+4.4}_{-8.3}$ seconds without any other extra constraints. Its relative precision is about $1.8\times10^{-3}$.

\acknowledgments
This work is co-sponsored by the NSFC of China (Grant Nos. 11333006, 11503076, and 11521303), and by the Chinese Academy of Sciences (Grant No. XDB09010202). The authors express their sincere thanks to NASA and the \kepler\ team for allowing them to work with and analyze the \kepler\ data making this work possible and also gratefully acknowledge the computing time granted by the Yunnan Observatories, and provided on the facilities at the Yunnan Observatories Supercomputing Platform. The \keplermission\ is funded by NASA's Science Mission Directorate. In addition, the authors are cordially grateful to an anonymous referee for instructive advice and productive suggestions.

\end{document}